\documentclass[twocolumn,
showpacs,groupedaddress,
preprintnumbers,aps,prd]{revtex4}
\usepackage{amsmath,amssymb, amsthm,mathrsfs}
\def\be{\begin{equation}}
\def\ee{\end{equation}}
\def\bea{\begin{eqnarray}}
\def\eea{\end{eqnarray}}
\def\ba{\begin{array}}
\def\ea{\end{array}}
\newcommand{\bes}{\begin{subequations}}
\newcommand{\ees}{\end{subequations}}
\def\nn{\nonumber}
\def\p{\partial}

\begin{document}

\preprint{hep-th/0701235 v3}
\title{Hawking radiation from the dilatonic black holes via anomalies}

\author{Qing-Quan Jiang$^{1)}$\footnote{E-mail address: jiangqingqua@126.com}}
\author{Shuang-Qing Wu$^{2)}$\footnote{E-mail address: sqwu@phy.ccnu.edu.cn}}
\author{Xu Cai$^{1)}$\footnote{E-mail address: xcai@mail.ccnu.edu.cn}}

\affiliation{$^{1)}$Institute of Particle Physics, Central China Normal University,
Wuhan, Hubei 430079, People's Republic of China \\
$^{2)}$College of Physical Science and Technology, Central China Normal University,
Wuhan, Hubei 430079, People's Republic of China}


\begin{abstract}
Recently, Hawking radiation from a Schwarzschild-type black hole via gravitational anomaly at
the horizon has been derived by Robinson and Wilczek. Their result shows that, in order to demand
general coordinate covariance at the quantum level to hold in the effective theory, the flux of
the energy momentum tensor required to cancel gravitational anomaly at the horizon of the black
hole, is exactly equal to that of $(1+1)$-dimensional blackbody radiation at the Hawking temperature.
In this paper, we attempt to apply the analysis to derive Hawking radiation from the event horizons
of static, spherically symmetric dilatonic black holes with arbitrarily coupling constant $\alpha$,
and that from the rotating Kaluza-Klein $(\alpha = \sqrt{3})$ as well as the Kerr-Sen ($\alpha = 1$)
black holes via an anomalous point of view. Our results support Robinson-Wilczek's opinion. In addition,
the properties of the obtained physical quantities near the extreme limit are qualitatively discussed.
\end{abstract}

\maketitle

\section{Introduction}\label{intro}

Since Stephen Hawking proved that a black hole can emit any kinds of particles from its event
horizon with a temperature proportional to its surface gravity, many derivations of Hawking
radiation had been proposed \cite{SWH,TRO,TRE,CPM} during the past thirty years. Among them,
some tied Hawking radiation to the contributions of the anomalies in different symmetries.
For example, Christensen and Fulling \cite{CF} showed that the anomaly in the conformal symmetry
could give strong constraints on the trace of the energy momentum tensor, and Hawking radiation
arose from the contribution of the trace anomaly of the energy momentum tensor \cite{BFS}. Imposing
the (1+1)-dimensional anomalous trace equation everywhere in the whole spacetime, a freely falling
observer would observe an outgoing flux, which is in quantitative agreement with Hawking's original
result. This seems to be a neat derivation of Hawking radiation, however, their observation was based
on several assumptions: Firstly, the background was in $(1+1)$ dimensions. Secondly, the fields were
massless. Finally, there were no back-scattering effect. Therefore, their derivation appeared to be
a rather special phenomenon.

Recently, Robinson and Wilczek \cite{RW} proposed a new derivation of Hawking radiation from a
Schwarzschild-type black hole via the anomaly in general coordinate symmetry at the horizon, namely
via gravitational anomaly. In their opinion, Hawking radiation is effectively treated as the required
flux of the energy momentum tensor to cancel gravitational anomaly and to restore general coordinate
covariance at the quantum level. In fact, the idea of this effective theory stems from the choice of
vacuum states. Normally, a static, spherically symmetric black hole without a positive cosmological
constant has a global Killing vector, but it is only timelike in the region outside the horizon of
the black hole. Thus, if one associates the positive energy with the occupation of modes of a positive
frequency, there would be a divergent energy momentum tensor due to a pile up of horizon-skimming modes
at the horizon since this time coordinate becomes mathematically ill-defined at the horizon. Replacing
the coordinates $(t, r)$ with the Kruskal extension $(U, V)$, then the final form of the spacetime
showed that $U \rightarrow U +U_0$ is an isometry along the past horizon $V = 0$, where $\xi \equiv
\p_U$ is a Killing vector.

As is well-known, there are various different ways to define the vacuum state in curved spaces. Unruh
\cite{US} introduced a quantum vacuum state, the so-called Unruh state or $\xi$-vacuum, for the static
spacetime by defining positive energy states as those that have positive frequency with respect to $\xi$
on the past horizon, and letting this vacuum propagate outward and forward in time. The vacuum state
defined in this way differs from the one obtained by defining positive energy with respect to the global
Killing vector $\eta$, which is called the Boulware state, or $\eta$-vacuum \cite{BS}. The $\eta$-vacuum
has divergences in its energy-momentum tensor arising from horizon-skimming modes, despite appearing empty
to static observers. This divergence renders the Boulware state an unphysical candidate vacuum. On the
other hand, the Unruh state is well behaved at the horizon with respect to the Kruskal coordinate $U$.
In addition, Unruh found that the $\xi$-vacuum is exactly a thermal ensemble of the Boulware states at
Hawking temperature $T = \kappa/(2\pi)$. If elevating the choice of a state to the level of a choice of
the theory, one expects that the effective theory should be formulated to exclude the offending modes
at the horizon, and describe a thermal flux, in exact agreement with the earlier results of Hawking
\cite{SWH}.

Since the effective field theory is formulated outside the horizon to integrate out the horizon-skimming
modes, it becomes chiral there, and suffers from gravitational anomaly because the numbers of the ingoing
and outgoing modes are no longer identical there. To demand general coordinate covariance at the quantum
level to hold in the effective theory, one must introduce a compensating flux of the energy momentum tensor
to cancel gravitational anomaly at the horizon of the black hole. As a result, the energy momentum tensor
flux, which is required to cancel gravitational anomaly and to restore general coordinate covariance at the
quantum level, is exactly equal to that of $(1+1)$-dimensional blackbody radiation at the Hawking temperature.

A key technique in the Robinson-Wilczek's method is to reduce the higher dimensional theory to the
two-dimensional case, in which gravitational anomaly appears as a chiral one. If each partial wave
of the higher-dimensional theory behaves, near the black hole horizon, as a $(1+1)$-dimensional blackbody
source at the Hawking temperature, one reproduces the standard calculation of Hawking radiation. Obviously
because the derivation of Hawking radiation via the anomalous point of view is only dependent on the information
of the horizon of the black hole, it is more universal than that from the trace anomaly. Following
this method, a lot of work that generalizes to various black hole cases appeared \cite{ISW1,ISW2,MS,HAE} in
recent months.

In the case of a charged black hole \cite{ISW1}, one should also consider the conservation of electric charge
beyond the energy conservation. In such case, the two-dimensional theory for each partial wave contains both
gauge and general coordinate symmetries, where the gauge symmetry arises from the gauge potential with respect
to the electric field of the original charged black hole. Near the horizon, when omitting the classically
irrelevant ingoing modes, the effective near-horizon quantum field becomes chiral, and contains the anomalies
related to these two symmetries, also known as gauge or gravitational anomalies. To demand gauge invariance
and general coordinate covariance at the quantum level, two kinds of compensating fluxes must be introduced
to cancel the corresponding gauge and gravitational anomalies at the horizon. Thee flux used to cancel the gauge
anomaly is called as the charge current flux, and the other one is the energy momentum tensor flux responsible
for cancelling the gravitational anomaly.

As far as the case of a rotating black hole is concerned, one should take into account the conservation
of angular momentum. After reduction to the ($1+1$)dimensions, each partial wave now contains a new $U(1)$
gauge symmetry besides the general coordinate symmetry. Although this $U(1)$ gauge symmetry originates
from the isometry along $\varphi$-direction, it can be viewed as arising from the $U(1)$ gauge potential
if written in terms of the angular velocity at the horizon of the rotating black hole, in which the $U(1)$
gauge charge $m$ is an azimuthal quantum number. This means that the two-dimensional reduction of the
rotating black hole can be equivalently treated as that of a charged particle in a two-dimensional charged
field. As such, the same procedure as that in the charged case can be applied to obtain the fluxes required
to cancel $U(1)$ gauge and gravitational anomalies at the horizon of the rotating black hole \cite{ISW2,MS,HAE}.
In order to demand gauge invariance and general coordinate covariance at the quantum level, the fluxes
derived in both cases via anomalies must be equal to that of $(1+1)$-dimensional blackbody radiation with
the Planck distribution, including chemical potentials for an electric charge $e$ or an azimuthal angular
momentum $m$ of the scalar fields radiated from the black holes.

Since the dilatonic black holes \cite{GM,GHS,HH,KK,KS} obtained from the low energy effective field theory
have qualitatively different properties from those in ordinary Einstein gravity, it is necessary and helpful
to study the thermal properties of these black holes via the anomalous point of view. In this paper, motivated
by Robinson-Wilczek's recent viewpoint, we study Hawking radiation from the static, spherically symmetric
dilatonic black holes with arbitrary coupling constant, and that from the rotating Kaluza-Klein and Kerr-Sen
dilatonic black holes \cite{KM} via gauge and gravitational anomalies. The result shows that, demanding gauge
invariance or general coordinate covariance at the quantum level, one can find each partial wave of the scalar
field to be in a state with a net charge current or energy momentum tensor flux, whose values are exactly equal
to that of $(1+1)$-dimensional blackbody radiation at the Hawking temperature with an appropriate chemical
potential.

The paper is outlined as follows. We begin with our studies in Sec. \ref{ghsbh} by applying the Robinson-Wilczek's
method of anomaly cancellation to study Hawking radiation from the static, spherically symmetric dilatonic black
holes with arbitrary coupling constant. Sec. \ref{kkbh} and Sec. \ref{ksbh} are, respectively, devoted to
investigating the cases of the rotating Kaluza-Klein and Kerr-Sen dilatonic black holes in four dimensions
via gauge and gravitational anomalies. Sec. \ref{dc} ends up with some discussions. In addition, special
properties of these dilatonic black holes in the extreme limit are also qualitatively discussed.

\section{Hawking radiation from the static, spherically symmetric dilatonic black hole}\label{ghsbh}

The action describing the dilaton field coupled to a $U(1)$ gauge field in (3+1) dimensions is
\be
S = \frac{1}{16\pi}\int d^4x\sqrt{-g}\Big[R -2(\nabla\phi)^2 -e^{-2\alpha\phi}F^2\Big] \, ,
\label{act}
\ee
where $\phi$ and $F$ are the dilaton field and the $U(1)$ gauge field, respectively, with a coupling constant
$\alpha$. From the action, the metric of the four-dimensional static, spherically symmetric dilatonic black hole
(with an arbitrary coupling constant $\alpha$), the dilaton field, and Maxwell field had been obtained as
\cite{GM,GHS,HH}
\be
\begin{aligned}
ds^2 &= -\frac{\Delta}{R^2}dt^2 +\frac{R^2}{\Delta}dr^2 +R^2\big(d\theta^2 +\sin^2\theta d\varphi^2\big) \, , \\
\phi &= \frac{\alpha}{1 +\alpha^2} \ln\Big(1 -\frac{r_-}{r}\Big) \, , \\
   F & = \frac{Q}{r^2} dt\wedge dr \, , \label{sdbh}
\end{aligned}
\ee
where
\bea
\Delta = (r -r_+)(r -r_-) \, , \qquad R = r\Big(1 -\frac{r_-}{r}\Big)^{\alpha^2/(1 +\alpha^2)} \, , \nn
\eea
in which the outer and inner horizons are, respectively, given by
\be
r_{\pm} = \frac{1 +\alpha^2}{1 \pm \alpha^2}\Big[M \pm \sqrt{M^2 -\big(1 -\alpha^2\big)Q^2}\Big] \, .
\ee
For $\alpha \neq 0$, $r = r_-$ is the curvature singularity. In the extreme limit (namely, $Q_{\max} =
\sqrt{1 +\alpha^2}M$), the inner horizon and the outer horizon coincide with each other, a naked singularity
then appears at $r = r_+$, and the area of black hole vanishes for $\alpha \neq 0$, the solution no longer
represents a black hole.

After performing the conformal transformation $\widetilde{g}_{\mu\nu} = e^{-2\alpha\phi}g_{\mu\nu}$, the line
element (\ref{sdbh}) for the black hole becomes
\be
d\widetilde{s}^2 = -f(r)dt^2 +\frac{1}{g(r)}dr^2 +r^2\big(d\theta^2 +\sin^2\theta d\varphi^2\big) \, ,
\ee
where
\be
f(r) = \frac{\Delta }{R^2}e^{-2\alpha\phi} \, , \qquad g(r) = \frac{\Delta}{R^2}e^{2\alpha\phi} \, .
\ee

Now, we focus on investigating Hawking radiation from the dilatonic black hole via gauge and gravitational anomalies
at the horizon. Near the black hole horizon, the effective radial potential for partial wave modes of the scalar
field vanishes exponentially fast if one introduces the tortoise coordinate defined by $r_* = \int (R^2/\Delta)dr$,
and performs the partial wave decomposition in terms of the spherical harmonics. Thus, the physics near the
horizon of the originally four-dimensional black hole can be effectively described by an infinite collection
of ($1+1$)-dimensional fields. In the effective two-dimensional reduction, the partial wave is in the background
of the metric and the gauge potential
\be
g_{tt} = -f(r) \, , \qquad g_{rr} = \frac{1}{g(r)} \, ,  \qquad A_t = -\frac{Q}{r} \, .
\ee
In addition, the background includes a dilaton field, but its contribution can be dropped since the effect
of dilaton field does not change the property of the $T_t^r$ component of the energy momentum tensor in the
static spacetime and accordingly the compensating flux is independent of the dilaton background \cite{ISW2}.

In the following, we firstly study the flux of the electric current and gauge anomaly at the horizon. In
the two-dimensional theory, the electric current is given by integrating the four-dimensional current over
a two-dimensional sphere. We define the effective field theory outside the horizon. In the region $r_+
+\epsilon \leq r$, there is no anomalies, so the current satisfies the conservation equation $\p_r
\big[\sqrt{-g_2}J_{(o)}^r\big] = 0$. Near the horizon $r_+ \leq r \leq r_+ +\epsilon$, when excluding
the near-horizon-skimming modes whose contributions would give a divergent electric current at the horizon,
the effective theory becomes chiral there, and contains the anomaly with respect to gauge symmetry, which
originates from the electric field of the charged black hole, also named as gauge anomaly. Thus, the current
there obeys the anomalous equation $\p_r \big[\sqrt{-g_2}J_{(H)}^r\big] = e^2\p_r A_t/(4\pi)$. What
needs to say is that, in the two-dimensions, the anomalous current actually satisfies $\nabla_{\mu} J^\mu
= -e^2\epsilon^{\mu\nu}\p_\mu A_\nu/(4\pi\sqrt{-g_2})$. However, one only needs to investigate the
electric current of the component $\mu = r$, since the temporal component of the electric current is
irrelevant for the Hawking radiation flux of the charge.

In the classical theory, gauge symmetry of the action is expressed by the covariant conservation of the charge
current under gauge transformations. However, in the quantized version, the modes' propagation along one light-like
direction makes the time coordinate ill-defined at the horizon for an observer outside the horizon. To remove
these offending modes, the current exhibits an anomaly in gauge symmetry. Under gauge transformations, if we
integrate the outgoing modes to obtain the effective action, then it changes as $-\delta W = \int dtdr \lambda
\nabla_{\mu} J^\mu $, where the current is written as $J^\mu = J_{(o)}^\mu \Theta_+(r) +J_{(H)}^\mu H(r)$, in
which $\Theta_+(r) = \Theta(r -r_+ -\epsilon)$ and $H(r) = 1 -\Theta_+(r)$ are the scalar step and top hat
functions, respectively. Thus the variation of the effective action under gauge transformations is given by
\be
\begin{aligned}
-\delta W &= \int dtdr \lambda \Big\{\p_r\Big[\frac{e^2}{4\pi}A_tH(r)\Big]
 +\Big[\frac{e^2}{4\pi}A_t \\
&\quad +\sqrt{-g_2}\big(J_{(o)}^r -J_{(H)}^r\big)\Big]\delta\big(r -r_+ -\epsilon\big)\Big\} \, ,
\label{eav}
\end{aligned}
\ee
where the ingoing modes are integrated out. To require the full quantum theory gauge invariance, the
quantum effect of the omitted ingoing modes should be taken into account. Since its contribution to
the total current flux is $-e^2A_tH(r)/(4\pi)$, the first term in Eq. (\ref{eav}) is cancelled by its
quantum effect \cite{ISW2}. To demand gauge invariance at the quantum level to hold in the effective
theory, the coefficient of the delta function should be nullified, which means that
\be
c_o = c_H -\frac{e^2}{4\pi}A_t(r_+) \, ,
\ee
where we have used the currents in both regions: $\sqrt{-g_2}J_{(o)}^r = c_o$, and $\sqrt{-g_2}J_{(H)}^r
= c_H +e^2\big[A_t(r) -A_t(r_+)\big]/(4\pi)$, in which $c_o$ and $c_H$ are the values of the current
at the infinity and the horizon, respectively. In order to fix the current flux, we impose a constraint
that demands the covariant current defined by $\widetilde{J}_{(H)}^r = J_{(H)}^r +e^2A_t(r_+)/(4\pi
\sqrt{-g_2})$ vanish at the horizon, which in fact, corresponds to regular requirement of the physical
quantities \cite{ISW2}. The electric current flux, required to cancel gauge anomaly and to demand
gauge invariance at the quantum level to hold in the effective theory, then reads off
\be
c_o = -\frac{e^2}{2\pi}A_t(r_+) \, .
\ee
In fact, the electric current flux is exactly equal to that of $(1+1)$-dimensional blackbody radiation with
the Planck distribution, including a chemical potential for an electric charge $e$.

Next, we will study the flux of the energy momentum tensor. To demand general coordinate covariance at the
quantum level, we expect that the flux is equal to that of Hawking radiation. When we exclude modes' propagation
along the light-like direction at the horizon, in the region $r_+ +\epsilon \leq r$ where the background
contains a gauge potential but has no anomaly, the energy momentum tensor satisfies the modified conservation
equation $\p_r \big[\sqrt{-g_2}T_{t(o)}^r\big] = c_o\p_r A_t$ deduced from the $\nu = t$ component of equation
$\nabla_{\mu}T_\nu^\mu = 0$, which is obeyed by the four-dimensional energy momentum tensor of the black hole. Near the
horizon, the offending modes has been integrated out, the effective theory becomes chiral, and contains gauge
and gravitational anomalies. Then the energy momentum tensor $T_t^r$ satisfies \cite{ISW2}
\bea
&& \p_r \big[\sqrt{-g_2}T_{t(H)}^r\big] = \sqrt{-g_2}J_{(H)}^r\p_r A_t \nn \\
&&\qquad\qquad\qquad\quad +A_t\p_r \big[\sqrt{-g_2}J_{(H)}^r\big] +\p_r N_t^r \, ,
\label{fae}
\eea
where, $N_t^r = \big(f_{,r}g_{,r} +gf_{,rr}\big)/(192\pi)$. It is suffice for our discussions to only
investigate the component $T_t^r$ since the anomaly is time-like. In the above equation, the second
term comes from gauge anomaly while the third one is from gravitational anomaly. In the effective theory,
since the energy momentum tensor combines contributions from both regions, that is, $T_{\nu}^\mu =
T_{\nu (H)}^\mu H(r) +T_{\nu (o)}^\mu \Theta_+(r)$, the effective action under general coordinate
transformation changes as
\bea
-\delta W &=& \int dtdr \lambda^t\Big\{c_o\p_rA_t +\p_r
 \Big[\Big(\frac{e^2}{4\pi}A_t^2 +N_t^r\Big)H(r)\Big] \nn \\
&& +\Big[\sqrt{-g_2}\big(T_{t(o)}^r -T_{t(H)}^r\big)
 +\frac{e^2}{4\pi}A_t^2 +N_t^r\Big] \nn \\
&&~ \times~ \delta\big(r -r_+ -\epsilon\big)\Big\} \, . 
\eea
where $\lambda^t$ is the general coordinate transformation parameter, and we have not incorporated the quantum
effect of the ingoing modes. The first term is the classical effect of the background electric field for constant
current flow, the second term is cancelled by the quantum effect of the ingoing modes whose contribution to the
energy momentum tensor flux is $-\big[N_t^r +e^2A_t^2/(4\pi)\big]$. To demand general coordinate symmetry of the
effective action at the quantum level, the energy momentum tensor fluxes in both regions satisfy
\be
a_o = a_H +\frac{e^2}{4\pi}A_t^2(r_+) -N_t^r(r_+) \, ,
\ee
where
\bea
\begin{aligned}
a_o &= \sqrt{-g_2}T_{(o)t}^r -c_oA_t(r) \, , \\
a_H &= \sqrt{-g_2}T_{t(H)}^r -\int_{r_+}^rd r\p_r\Big(c_oA_t +\frac{e^2A_t^2}{4\pi} +N_t^r\Big) \, , \nn
\end{aligned}
\eea
are the values of the energy flow at the infinity and the horizon, respectively.

Taking the form of the covariant energy momentum tensor as
\be
\sqrt{-g_2}\widetilde{T}_{t}^r = \sqrt{-g_2}T_{t}^r +\frac{g}{192\pi f}\Big(ff_{,rr}
-2f_{,r}^2\Big) \, ,
\ee
and further imposing the vanishing condition on it, which in fact corresponds to the regularity condition for
the energy momentum tensor at the future horizon \cite{ISW2}, one will find the flux of the energy momentum
tensor, required to demand general coordinate covariance at the quantum level, is given by
\be
a_o = \frac{e^2}{4\pi}A_t^2(r_+) +\frac{\pi}{12}T_+^2 \, ,
\label{emfc}
\ee
where
\be
\begin{aligned}
T_+ &= \frac{1}{4\pi}\sqrt{f_{,r}g_{,r}}\big|_{r = r_+} \\
    &= \frac{1}{4\pi r_+}\Big(1 -\frac{r_-}{r_+}\Big)^{(1 -\alpha^2)/(1 +\alpha^2)} \, ,
\end{aligned}
\ee
is the Hawking temperature of the black hole. The energy momentum tensor flux exactly agrees with that of
Hawking radiation from the black hole.

Finally let us study the electric current and energy momentum tensor fluxes of Hawking radiation in the case
of fermions. For the charged black hole, the Hawking distribution for fermions is given by $N_{\pm e}(\omega)
= 1/\big[\exp(\frac{\omega\mp e A_t(r_+)}{T_+}) +1\big]$, with which the electric current and energy momentum
tensor fluxes have the completely equivalent forms to those obtained by the cancellation conditions of gauge
and gravitational anomalies and the regularity requirement at the horizon \cite{ISW2,MS}. So, Hawking radiation
can be derived from the anomalous point of view.

The spherically symmetric dilatonic black holes are derived from the low-energy string theory, and the background
has a dilaton field coupled to a gauge field. The solutions have some interesting thermodynamical properties
especially in the extreme case \cite{HH,DBH}, which are not found in the conventional black hole. The temperature
of the spherically symmetric dilatonic black holes in the extreme limit depends drastically on the dilaton
coupling constant $\alpha$. When the charge of the black hole approaches to the maximum value $Q_{\max}
= \sqrt{1 +\alpha^2}M$, the temperature of the black hole with $\alpha < 1$ vanishes as it does in the
Reissner-Nordstr\"{o}m black hole case. If $\alpha > 1$, it diverges in the extreme limit. When moving to
the case $\alpha = 1$, it has the nonzero finite value $1/(8\pi M)$. Besides, the energy momentum tensor
flux for each value of the dilaton coupling constant $\alpha$ coincides at $Q = 0$, since the black hole
solution with any $\alpha$ is identically the Schwarzschild spacetime for $Q = 0$ \cite{KM}; the difference
becomes large as the charge increases. When the charge of the black hole is fixed, the charge current flux
is eliminated from the anomalous flux. The energy momentum tensor flux in Eq. (\ref{emfc}) is then nonzero
but finite at $\alpha = 1$ near the extreme limit. However, in the extreme case, it vanishes at $\alpha <
1$, while the black hole with $\alpha > 1$ radiates a large amount of energy. So, the static, spherically
symmetric dilatonic black holes with the coupling constant $\alpha > 1$ evolve rapidly into a naked
singularity near the extreme case. In that case, we have to resort to the full quantum theory to
properly study such problems.

\section{Hawking radiation from the rotating Kaluza-Klein dilatonic black hole}\label{kkbh}

In the rotating case, the Kaluza-Klein black hole is an exact solution to the action (\ref{act}) with the
coupling constant $\alpha = \sqrt{3}$. It is derived by a dimensional reduction of the boosted five-dimensional
Kerr solution to four dimensions. The metric is given by \cite{KK,KM}
\bea
ds^2 &=& -\frac{\Delta -a^2\sin^2\theta}{B\Sigma}dt^2 -2a\sin^2\theta\frac{Z}{B\sqrt{1 -\nu^2}}dtd\varphi \nn \\
&& +\Big[B(r^2+a^2) +a^2\sin^2\theta \frac{Z}{B}\Big]\sin^2\theta d\varphi^2 \\
&& +\frac{B\Sigma}{\Delta}dr^2 +B\Sigma d\theta^2 \, , \nn
\eea
where
\be
\begin{aligned}
&\Delta = r^2 -2\mu r +a^2 \, , \qquad \Sigma = r^2 +a^2\cos^2\theta \, , \\
& Z = \frac{2\mu r}{\Sigma} \, , \qquad\qquad\qquad B = \sqrt{1 +\frac{\nu^2Z}{1 -\nu^2}} \, .
\end{aligned}
\ee
The dilaton field is $\phi = -\big(\sqrt{3}/2\big)\ln B$, and the potential corresponding to the gauge field is
\be
A_t = \frac{\nu Z}{2(1 -\nu^2)B^2} \, , \qquad
A_{\varphi} = -\frac{a\nu Z\sin^2\theta}{2\sqrt{1 -\nu^2}B^2} \, .
\ee
The physical mass $M$, the charge $Q$ and the angular momentum $J$ are expressed, by the boost parameter $\nu$,
mass parameter $\mu$ and specific angular momentum $a$, as
\be
\begin{gathered}
M = \mu \Big[1 +\frac{\nu^2}{2(1 -\nu^2)}\Big] \, , \qquad\qquad \\
Q = \frac{\mu\nu}{1 -\nu^2} \, , \qquad J = \frac{\mu a}{\sqrt{1 -\nu^2}} \, ,
\end{gathered}
\ee
respectively. When $Q = 0$, the black hole is reduced to Kerr black hole. The $J = 0$ case corresponds to the
non-rotating black hole with the same coupling constant $\alpha=\sqrt{3}$. The event horizon of the black hole
is given by $r_+ = \mu +\sqrt{\mu^2 -a^2}$, and the nonsingular inner horizon is located at $r_- = \mu -\sqrt{\mu^2
-a^2}$. The condition that the black hole has a regular horizon is $\mu^2 \geq a^2$. It should be
noticed that the solution, in the extreme limit ($|Q| = 2M$ and $J = 0$), is no longer representing a black
hole because a naked singularity appears \cite{KM}. This is quite different from the Kerr-Newman ($\alpha = 0$)
black hole.

Now, we study Hawking radiation from the rotating Kaluza-Klein black hole via the anomalous point of view.
Near the horizon, performing the partial wave decomposition of the scalar field in term of the spherical harmonics
$\varphi = \sum_{l, m} \varphi_{lm}(t, r)Y_{lm}(\theta, \phi)$ [Strictly speaking, the angular part of the separated
scalar field equation can be transformed into a form of confluent Heun equation \cite{WC}, but near the horizon it
approaches to the spherical harmonic since the near-horizon geometry has a topology of 2-sphere.] and transforming
to the tortoise coordinate defined by $dr_*/dr = B\big|_{\theta = 0}(r^2+a^2)/\Delta \equiv 1/f(r)$, one can easily
observe that the effective two-dimensional theory, decided by the following metric and gauge field
\be
\begin{aligned}
ds^2 &= -f(r)dt^2 +f^{-1}(r)dr^2 \, , \\
\mathcal{A}_t &= -\frac{Qr e(1 -\nu^2)}{r^2 +a^2} -\frac{am\sqrt{1 -\nu^2}}{r^2 +a^2} \, ,
\end{aligned}
\ee
is capable of describing the physics near the horizon of the higher-dimensional black hole. Here, we still omit
the dilaton background due to the reduced static space-time. In the gauge field background, $e$ and $m$ are the
electric charge and azimuthal angular quantum number of the scalar field, respectively. The first term stems from
the electric field of the charged black hole, and the second one is the induced gauge potential associated with
the black hole's axisymmetry. So, the two-dimensional reduction includes general coordinate symmetry and two gauge
symmetries. When omitting the classically irrelevant ingoing modes near the horizon, the effective action becomes
anomalous with respect to these symmetries. In order to relieve the conflict between a symmetry of the classical
action and the procedure of quantization, one must introduce the corresponding fluxes to cancel these anomalies.

As before, the first thing is to study gauge current flux and gauge anomalies. The effective field theory is
defined outside the horizon. Near the horizon $r_+ \leq r \leq r_+ +\epsilon$, when excluding the horizon-skimming
modes, the current $\widetilde{J}$ derived from the electric current in the original charged black hole, becomes
anomalous and satisfies the anomalous equation \cite{ISW2} as $\p_r \widetilde{J}^r_{(H)} = e\p_r
\mathcal{A}_t/(4\pi)$. [In fact, the gauge potential corresponding to the electric charge $e$ in the two-dimensional
reduction is given by $\mathcal{\widetilde{A}}_t = \mathcal{A}_t/e$, and the current should satisfy $\p_r
\widetilde{J}^r_{(H)} = e^2\p_r \mathcal{\widetilde{A}}_t/(4\pi)$.] In the other region, there is no
anomalies and the gauge current satisfies the conservation equation $\p_r \widetilde{J}^r_{(o)} = 0$.
Under gauge transformations, to demand the effective action gauge invariance at the quantum level, the
compensating flux of the current, is then written as
\be
d_o = -\frac{e}{2\pi}\mathcal{A}_t(r_+) \, ,
\ee
where we have imposed the condition that the covariant gauge current, which is expressed by $ \widehat{J}^r_{(H)}
= \widetilde{J}^r_{(H)} +e\mathcal{A}_t(r_+)/(4\pi)$, vanishes at the future horizon \cite{ISW2,MS}. This result
is in agreement with the electric current flux of Hawking radiation.

In addition to the energy and charge conservation, one should also take into account the conservation of angular
momentum in the rotating Kaluza-Klein dilatonic black hole. In two-dimensional theory, the axisymmetry of the
rotating black hole can now be interpreted as $U(1)$ gauge symmetry for the two-dimensional scalar field. There
is an induced $U(1)$ gauge potential corresponding to this $U(1)$ gauge symmetry, whose gauge charge $m$ is an
azimuthal angular momentum quantum number. When the effective field theory is defined outside the horizon, this
$U(1)$ gauge current $\widehat{J}$, which is deduced from the $(r, \varphi)$-component of the four-dimensional
energy momentum tensor, is composed of contributions from two regions. In the region $r_+ +\epsilon \leq r$, there is no
anomalies, and the gauge current satisfies the conservation equation $\p_r \widehat{J}^r_{(o)} = 0$. But
near the horizon, it exhibits an anomaly with respect to $U(1)$ gauge symmetry since the offending modes are
removed and the effective theory becomes chiral here, and satisfies the anomalous equation $\p_r
\widehat{J}^r_{(H)} = m\p_r \mathcal{A}_t/(4\pi)$, where the azimuthal quantum number $m$ is treated as
the $U(1)$ gauge charge. Under gauge transformations, the $U(1)$ gauge current flux, required to cancel $U(1)$
gauge anomaly, reads off
\be
f_o = -\frac{m}{2\pi}\mathcal{A}_t(r_+) \, .
\ee
This factually corresponds to the angular momentum flux of Hawking radiation.

We now investigate the flux of the energy momentum tensor. In classical theory, the symmetry of the action
under general coordinate transformations requires the covariant conservation of energy momentum tensor.
But as mentioned before, there is a divergent energy momentum tensor near the horizon due to a pile up of
the horizon-skimming modes. Suppose that the effective field theory is formulated to exclude the offending
modes, it no longer has a divergent energy momentum tensor near the horizon, but contains an anomaly
with respect to general coordinate symmetry and takes the form of the nonconservation of energy momentum
tensor. Adopting this picture, one can reformulate the effective field theory outside the horizon. In
the region $r_+ +\epsilon \leq r$, it contains all the same modes as the fundamental theory does, so there
is no anomaly. The energy momentum tensor, under an effective background gauge field, satisfies the modified
conservation equation $\p_rT^r_{t(o)} = \mathcal{J}^r_{(o)}\p_r\mathcal{A}_t$. However, it
becomes anomalous near the horizon since the numbers of the outgoing modes and ingoing modes are no longer
matched with each other when omitting the classically irrelevant ingoing modes. The energy momentum
tensor near the horizon satisfies the anomalous equation
\be
\p_rT^r_{t(H)} = \mathcal{J}^r_{(H)}\p_r\mathcal{A}_t
+\mathcal{A}_t\p_r\mathcal{J}^r_{(H)} +\p_rN^r_t \, ,
\ee
where the current is defined by $\mathcal{J}^r \equiv \widetilde{J}^r/e = \widehat{J}^r/m$, and $N^r_t =
\big(f_{,r}^2 +ff_{,rr}\big)/(192\pi)$. Imposing the boundary conditions that the covariant energy momentum
tensor, defined by $\widehat{T}^r_t = T^r_t +\big(ff_{,rr} -2f_{,r}^2\big)/(192\pi)$, vanishes at the horizon
and there is no ingoing modes at radial infinity, the covariance of the effective theory at the quantum level,
under general coordinate transformations, requires a compensating flux of the energy momentum tensor
\be
g_o = \frac{1}{4\pi}\mathcal{A}^2_t(r_+) +\frac{\pi}{12}T^2_+ \, ,
\ee
where
\be
\begin{aligned}
T_+ = \frac{1}{4\pi}\p_r{f}\big|_{r = r_+}
= \frac{\sqrt{1 -\nu^2}\sqrt{\mu^2 -a^2}}{2\pi(r^2_+ +a^2)} \, ,
\end{aligned}
\ee
is the Hawking temperature of the black hole. In fact, the energy momentum tensor flux is exactly equal to
that of Hawking radiation.

As for the fermionic case, the Planck distribution (including the chemical potentials for an azimuthal
angular quantum number $m$ and an electric charge $e$), the electric potential and the angular velocity
at the black hole horizon are, respectively, given by
\bea
\begin{gathered}
N_{\pm e, \pm m}(\omega) = 1/\big[\exp{(\omega\mp e\Phi_+ \mp m\Omega_+)} +1\big] \, , \nn \\
\Phi_+ = \frac{Qr_+ (1 -\nu^2)}{r^2_+ +a^2} = \frac{\nu}{2} \, , \qquad
\Omega_+ = \frac{a\sqrt{1 -\nu^2}}{r^2_+ +a^2} \, ,
\end{gathered}
\eea
the electric current, angular momentum, and energy momentum tensor fluxes of Hawking radiation have the forms
equivalent to those required to cancel gauge and gravitational anomalies and to demand gauge invariance and
general coordinate covariance at the quantum level to hold in the effective theory \cite{ISW2}. Thus, Hawking
radiation can be effectively described from the anomalous point of view.

Now, we study the properties of the derived physical quantities in the extreme case. Obviously, the
temperature of the rotating dilatonic black hole vanishes in the extreme limit $\mu = a$. In addition, when
taken the limit $Q \to Q_{\max} = 2M$, keeping the black hole extreme with $J \neq 0$ (whereas $J \to 0$ in
the extreme), the value of the temperature is still zero, which is different from the nonrotating dilatonic
black holes, where the temperature diverges in the limit $Q \to Q_{\max} = 2M$. Thus the temperature is
discontinuous at $Q = Q_{\max}$; this is due to the fact that there is a naked singularity for the
Klauza-Klein black hole when reaching to the extreme case. Moreover, the angular velocity $\Omega$,
under the limit $Q \to Q_{\max}$, diverges at the horizon, but the angular momentum $J$ vanishes.
This is because the horizon shrinks to zero size as the angular momentum decreases. The solution with $Q
\to Q_{\max}$ then represents a rapidly spinning black hole \cite{KM}.

Subsequently, we will vary the charge to see how the angular momentum and energy momentum tensor fluxes
change in the maximally charged limit. As for the rotating Kaluza-Klein dilatonic black hole, the temperature
is zero in the extreme case, and the thermal emission vanishes. However, in the maximally charged limit
$Q \to Q_{\max}$, while the angular momentum itself is small, the angular velocity of the black hole is
very large and its superradiance effect becomes important. The angular momentum and energy momentum
tensor fluxes increase rapidly till the black hole approaches to the maximally charge state, where
the horizon radius is smaller than the Planck scale. However a better understanding of physics at the
Planck scale is a prerequisite; this is especially necessary for studying black holes at the maximally
charge state, which have to rely on the full quantum theory.

\section{Hawking radiation from the rotating Kerr-Sen dilatonic black hole}\label{ksbh}

The Kerr-Sen black hole \cite{KS} is a solution to the low-energy effective action in heterotic string
theory. The action containing a three-form (axion) field $H$ and a dilaton field $\phi$ coupled to a
$U(1)$ gauge field $F$ reads off \cite{KS,KM}
\be
S = \frac{1}{16\pi}\int d^4x\sqrt{-g}\Big[R -2{(\nabla\phi)}^2
-e^{-2 \phi}F^2 -\frac{1}{12}e^{-4\phi}H^2\Big] \, .
\ee

From the uncharged Kerr solution, Sen \cite{KS} adopted the solution generating technique to obtain a
new rotating one, which is given by
\bea
ds^2 &=& -\frac{\Delta -a^2\sin^2\theta}{\Sigma}dt^2
-\frac{4\mu ra\cosh^2\beta\sin^2\theta}{\Sigma}dtd\varphi \nn \\
&& +\frac{\Lambda \sin^2\theta}{\Sigma}d\varphi^2
+\frac{\Sigma}{\Delta}dr^2 +\Sigma d\theta^2 \, ,
\eea
where
\be
\begin{aligned}
\Sigma &\equiv r^2 +a^2\cos^2\theta +2\mu r\sinh^2\beta \, , \\
\Delta &\equiv r^2 -2\mu r +a^2 \, , \\
\Lambda &\equiv (r^2 +a^2)(r^2 +a^2\cos^2\theta) +2\mu ra^2\sin^2\theta \\
&\quad +4\mu r(r^2 +a^2)\sinh^2\beta +4\mu^2r^2\sinh^4\beta \, .
\end{aligned}
\ee
The dilaton field, axion field and gauge potential corresponding to the gauge field are, respectively,
given by
\bea
&&\phi = \frac{-1}{2}\ln\frac{\Sigma}{r^2 +a^2\cos^2\theta} \, , ~
B_{t\varphi} = 2a\sin^2\theta\frac{\mu r\sinh^2\beta}{\Sigma} \, , \nn \\
&&A_t = \frac{\mu r \sinh 2\beta}{\sqrt{2}\Sigma} \, , \quad
A_{\varphi} = -a\sin^2\theta\frac{\mu r\sinh 2\beta}{\sqrt{2}\Sigma} \, .
\eea
The mass $M$, the charge $Q$, and the angular momentum $J$ are given by the parameters $\mu$, $\beta$
and $a$ as
\be
\begin{aligned}
M &= \frac{\mu}{2}(1 +\cosh 2\beta) \, , \\
Q &= \frac{\mu}{\sqrt{2}}\sinh2\beta \, , \quad J = Ma \, .
\end{aligned}
\ee
The horizon radius is determined as $r_+ = \mu +\sqrt{\mu^2 -a^2}$. The condition for the solution to
be a black hole is $\mu \geq |a|$, which can be rewritten as
\be
\big|J\big|\leq M^2 -Q^2/2 \, .
\ee
In the extreme limit $|Q| = \sqrt{2}M$ and $J=0$, a naked singularity appears, so the solution is no
longer a black hole \cite{KM}.

Following the reduction procedure adopted above, the physics near the horizon in higher dimensional black
hole can be effectively described by a two-dimensional theory. In the present case, each partial wave in
the near-horizon limit behaves as an independent two dimensional scalar field in the background of the
dilaton field whose contributions are omitted due to the static background. The two-dimensional metric
and the gauge field are
\be
\begin{aligned}
ds^2 &= -f(r)dt^2 +f^{-1}(r)dr^2 \, , \\
\mathcal{\widetilde{A}}_t &= -\frac{Qre}{(r^2 +a^2)\cosh^2\beta}
 -\frac{2\mu ar}{{(r^2 +a^2)}^2\cosh^2\beta} \, .
\end{aligned}
\ee
where, $f(r) = \Delta/(r^2 +a^2 +2\mu r\sinh^2\beta)$. Apparently, the two-dimensional theory for each partial
wave has two gauge symmetries. If an effective field theory is formulated outside the horizon to integrate out
the horizon-skimming modes, the electric current, angular momentum, and energy momentum tensor become anomalous
with respect to gauge and general coordinate symmetries near the horizon.  To demand gauge invariance and general
coordinate covariance at the quantum level to hold in the effective theory, one must introduce the compensating
electric current, angular momentum, and energy momentum tensor fluxes to cancel these anomalies at the horizon,
and they are precisely equal to those of Hawking radiation.

As usual, we will first study the electric current flux and its gauge anomaly. The effective theory is still
formulated outside the horizon to integrate out the horizon-skimming modes. Near the horizon, it becomes
chiral and contains gauge anomaly. Under gauge transformations, the electric current exhibits an anomaly
with respect to gauge symmetry, whose consistent form satisfies the anomalous equation $\p_r j_{(H)}^r
= e \p_r \mathcal{\widetilde{A}}_t/(4\pi)$. In the other region where the effective theory contains
all modes, the current satisfies the conservation equation $\p_r j_{(o)}^r = 0$, in which $j$ is
derived from the electric current in the four-dimensional spacetime. Under gauge transformations, to
demand the effective action gauge invariance at the quantum level, the electric current describes a
flux given by
\be
h_o = -\frac{e}{2\pi}\mathcal{\widetilde{A}}_t(r_+) \, ,
\ee
where we have already imposed the condition that the covariant form of the current vanishes at the horizon
\cite{ISW2,MS}. This electric current flux, derived from gauge anomaly cancellation and the  regularity
requirement at the horizon, is exactly equal to that of Hawking radiation.

Similarly, we can derive the flux of the $U(1)$ gauge current. The gauge current $\widetilde{j}$, deduced from
the component of the four-dimensional energy momentum tensor $T_\varphi^r$, becomes anomalous with respect to
$U(1)$ gauge symmetry when excluding the offending modes near the horizon. In the two-dimensional theory, the
$U(1)$ gauge current near the horizon satisfies the anomalous equation $\p_r \widetilde{j}_{(H)}^r = m
\p_r \mathcal{\widetilde{A}}_t/(4\pi)$. However, it is conserved $\p_r \widetilde{j}_{(o)}^r = 0$
in the region $r_+ +\epsilon \leq r$. The $U(1)$ gauge current flux, required to cancel $U(1)$ gauge anomaly,
is then given by
\be
i_o = -\frac{m }{2\pi}\mathcal{\widetilde{A}}_t(r_+) \, ,
\ee
where we have assumed the vanishing condition of the covariant $U(1)$ gauge current at the horizon. This
angular momentum flux exactly corresponds to that of Hawking radiation.

The above two gauge anomalies stem from the destruction of gauge symmetries associated with the electric field
and the axisymmetry of the rotating charged black hole. Besides these, gravitational anomaly in general coordinate
symmetry appears in the two-dimensional background when we omit the quantum effect of the ingoing modes. This is
formally reflected by the nonconservation of the energy momentum tensor. In the following, the energy momentum
tensor flux, required to restore general coordinate covariance at the quantum level, is derived from the anomalous
point of view. We expect that the value of the energy momentum tensor flux is equal to that of Hawking radiation
similar to the cases of the two gauge current fluxes. Near the horizon, the energy momentum tensor satisfies the
anomalous equation (\ref{fae}) with the replacements $N_t^r = \big(f_{,r}^2 +ff_{,rr}\big)/(192\pi)$ and $A_t =
\mathcal{\widetilde{A}}_t$. In the other region, there is no anomaly, but it contains a gauge field, the energy
momentum tensor satisfies the modified conservation equation $\p_rT_{t(o)}^r = \mathcal{J}_{(o)}^r\p_r
\mathcal{\widetilde{A}}_t$. In both regions, the current is defined by $\mathcal{J}^r\equiv j^r/e = \widetilde{j}^r/m$.
Imposing the condition that the covariant energy momentum tensor vanishes at the horizon, the energy momentum tensor
flux, required to demand the effective action general coordinate covariance at the quantum level, is expressed by
\be
k_o = \frac{1}{4\pi}\mathcal{\widetilde{A}}^2_t(r_+) +\frac{\pi}{12}T^2_+ \, ,
\ee
where $T_+$ is the Hawking temperature of the black hole, whose obvious expression is given by
\be
T_+ = \frac{1}{4\pi}\p_rf(r)\big|_{r = r_+}
= \frac{\sqrt{\mu^2 -a^2}}{2\pi (r_+^2 +a^2)\cosh^2\beta} \, .
\ee
This fits in with the energy momentum tensor flux of blackbody radiation at the Hawking temperature with Planck
distribution, including chemical potentials for an electric charge $e$ and an azimuthal quantum number $m$.

If we introduce the Planck distribution for fermions in the Kerr-Sen black hole as that in previous section,
and replace the electric potential by $\Phi_+ = Q/(2M) = Qr_+/[(r^2_+ +a^2)\cosh^2\beta]$, and the angular
velocity by $\Omega_+ = a/[(r^2_+ +a^2)\cosh^2\beta]$, the electric current, angular momentum, and energy
momentum tensor fluxes of Hawking radiation still have the equivalent forms to those derived from anomaly
cancellation condition and the regularity requirement at the horizon.

In the extreme limit $\mu = a$, the temperature and the angular velocity are no longer zero or divergent but
approach to finite values, however they still discontinue at the maximally charged limit $Q = Q_{\max} = \sqrt{2}M$,
since there exists a naked singularity for the extreme black hole. The angular momentum and energy momentum tensor
fluxes of the Kerr-Sen black hole increase rapidly, but slower than those of the Kaluza-Klein black hole, towards
the maximally charged limit. The final state is the presence of a naked singularity and the surface of the black
hole vanishes, which means that we have to deal with a horizon radius smaller that the Planck scale. That needs
to be investigated in the full quantum theory.

\section{Discussions and Conclusions}\label{dc}

In summary, we have applied the method of cancellation of anomaly to derive Hawking radiation of various
dilatonic black holes. After reductions from the four-dimension theory, the effective $(1+1)$-dimensional
field theory is formulated outside the horizon, which is largely based upon the choice of the vacuum state,
namely the Unruh vacuum rather than the Boulware state. Excluding the offending modes at the horizon results
in a breakdown at the quantum level of gauge invariance and general coordinate covariance. To demand these
symmetries to hold in the effective theory, the compensating fluxes can be obtained by the anomaly cancellation
and the regularity requirement at the horizon. In all cases, we find that the charge and energy momentum tensor
fluxes, required to cancel the gauge and gravitational anomalies, are exactly in agreement with those of
$(1+1)$-dimensional blackbody radiation at the Hawking temperature with Planck distribution including
appropriate chemical potentials.

It should be emphasized that in the cases of rotating and charged dilatonic black holes, the Robinson-Wilczek's
proposal to derive Hawking radiation can be also verified in the dragging coordinate system. In a rotating
spacetime, the matter field in the ergosphere near the horizon must be dragged by the gravitational field
with an azimuthal angular momentum because there exists a frame dragging effect of the coordinate. In the
dragging coordinate system, the matter field is co-rotating with the black hole, so the $U(1)$ gauge symmetry
induced from the azimuthal symmetry no longer needs to be incorporated in the two-dimensional theory for
each partial wave, and the corresponding gauge anomaly is excluded from this effective theory. To demand
gauge invariance at the quantum level to hold in the original underlying theory, the compensating flux of
the gauge current is only attributed to the contributions of gauge anomaly originated from the electric
field of the original charged black hole. The fluxes determined by the anomaly cancellations and the regularity
requirements at the horizon still have the completely equivalent forms to those of blackbody radiation
with Planck distribution function in the dragging coordinate.

In addition, the derivation of Hawking radiation via anomalies is based upon quantum field theory on the fixed
background spacetime without considering the fluctuation of spacetime geometry. With the evaporation, the mass,
the charge and the angular momentum of the realistic black hole must diminish and the background geometry
must vary accordingly. Thus a bigger challenge is to incorporate the self-gravitation correction into
the framework. Finally, in the maximally charged limit, the area of the charged dilatonic black hole vanishes,
but the fluxes of the angular momentum and energy momentum tensor are still very large. To properly study
Hawking radiation in these cases, the back reaction of the quantum effects must be taken into account and
one has to resort to the full quantum theory.

{\bf Acknowledgments}: ~S.-Q.Wu was partially supported by the Natural Science Foundation of China under Grant
No. 10675051 and by a starting fund from Central China Normal University. X.Cai was supported in part by the
NSFC of China (No. 70571027, No. 10635020) and by the Ministry of Education of China under Grant No. 306022.

\end{document}